\documentclass[pra,aps,superscriptaddress,nofootinbib,twocolumn]{revtex4-1}
\usepackage{mathrsfs}
\usepackage{amsfonts}
\usepackage{amssymb}
\usepackage{amsmath}
\usepackage{graphicx}
\usepackage{qcircuit}
\usepackage{algorithm}
\usepackage[noend]{algpseudocode}
\usepackage{algpseudocode}
\usepackage{subfigure,dcolumn}
\usepackage{graphicx}
\usepackage{mathtools}
\usepackage{braket}
\usepackage{subfigure}
\usepackage{epstopdf}
\usepackage[usenames,dvipsnames]{color}
\usepackage[colorlinks=true,citecolor=blue,linkcolor=magenta]{hyperref}
\usepackage{ulem}
\usepackage{lmodern}
\usepackage{amsthm}

\newcommand{\norm}[1]{\Vert #1 \Vert}
\newcommand{\normLR}[1]{\left\Vert #1 \right\Vert}
\newcommand{\abs}[1]{\vert #1 \vert}

\newcommand{\ketbra}[2]{\vert #1 \rangle \langle #2 \vert}

\newcommand{\NOTE}[1]{{\color{red} \it [[[#1]]]}}
\usepackage{listings}
\usepackage{algorithm}  
\usepackage{algpseudocode}  
\usepackage{amsmath}  

\begin{document}

\title{Quantum algorithms for optimal effective theory of many-body systems}

\author{Yongdan Yang}
\affiliation{Graduate School of China Academy of Engineering Physics, Beijing 100193, China}

\author{Zongkang Zhang}
\affiliation{Graduate School of China Academy of Engineering Physics, Beijing 100193, China}

\author{Xiaosi Xu}
\affiliation{Graduate School of China Academy of Engineering Physics, Beijing 100193, China}

\author{Bing-Nan Lu}
\email{bnlv@gscaep.ac.cn}
\affiliation{Graduate School of China Academy of Engineering Physics, Beijing 100193, China}

\author{Ying Li}
\email{yli@gscaep.ac.cn}
\affiliation{Graduate School of China Academy of Engineering Physics, Beijing 100193, China}

\begin{abstract}
A common situation in quantum many-body physics is that the underlying theories are known but too complicated to solve efficiently. 
In such cases one usually builds simpler effective theories as low-energy or large-scale alternatives to the original theories.
Here the central tasks are finding the optimal effective theories and proving their equivalence to the original theories. Recently quantum computing has shown the potential of solving quantum many-body systems by exploiting its inherent parallelism. It is thus an interesting topic to discuss the emergence of effective theories and design efficient tools for finding them based on the results from quantum computing. As the first step towards this direction, in this paper, we propose two approaches that apply quantum computing to find the optimal effective theory of a quantum many-body system given its full Hamiltonian. The first algorithm searches the space of effective Hamiltonians by quantum phase estimation and amplitude amplification. The second algorithm is based on a variational approach that is promising for near-future applications.
\end{abstract}

\keywords{Effective Model \quad Quantum computation .}

\maketitle

\section{Introduction}

    Effective theories are powerful tools in exploring quantum many-body systems. For many large systems, even though the exact Hamiltonian is known, it is difficult to solve the full quantum many-body problem. The challenge stems from the exponentially increasing Hilbert space as the system size increases. Fortunately, in many cases only a few relevant degrees of freedom participate in producing large-scale phenomena. This enables us to design effective theories which are usually much simpler and easier to solve than the original theories. More importantly, these theories isolate the most relevant symmetries and degrees of freedom and provide valuable insights into the underlying physics.
    One example is the quantum chromodynamics (QCD) for strong interaction, which is successful in describing quarks and gluons but still not directly solvable for a finite nucleus. 
    For nuclear physics applications, we are mainly interested in the low-energy regime where the quarks and gluons are confined and frozen in nucleons and light mesons.
    In these cases we can integrate out the irrelevant degrees of freedom and end up with an effective theory for nucleons and mesons, i.e., the chiral effective field theory ($\chi$EFT)~\cite{bib:1,bib:2}.
    As the most successful low-energy alternative to the QCD, the $\chi$EFT has been extensively used in solving nuclei as heavy as $^{208}$Pb~\cite{bib:3,bib:4,bib:5}, nuclear matter at zero~\cite{bib:6} or finite temperatures~\cite{bib:7} and low-energy nuclear reactions~\cite{bib:8,bib:9}.
    
In the early years, effective theories were usually established by trial and error, which sometimes relied on acute physics intuitions.
 Later it was realized that their form could be mostly constrained by a class of principles such as symmetry, analyticity and renormalizability.
In cases where the underlying microscopic theories are already known, the Wilsonian renormalization group (RG) method~\cite{bib:10,bib:11} provides a systematically improvable framework for eliminating irrelevant degrees of freedom and obtaining low-energy effective theories.
However, the Wilsonian RG is difficult to apply in many important cases, especially when the interaction is non-perturbative, such as the QCD at low energies.
Today most effective theories are still built from scratch under the aforementioned general principles and parametrized by matching to the experiments or underlying theories.
The validity of an effective theory is unknown until it is solved and a comparison is made between its predictions and experiments.
Unfortunately, effective theories are still very difficult to solve in some cases, in addition to the above restrictions.
A famous example is that the two-dimensional Ising model for ferromagnetic systems was thought to exhibit no phase transition for a long time, until Peierls published his argument~\cite{bib:12}. Today, even with the most advanced supercomputers and algorithms, many effective theories are still not solvable, which prohibits us from essentially understanding the relevant physics.  

Quantum computing is a fundamentally different paradigm for performing calculations compared with its classical counterpart. Because of its inherent property, i.e., quantum parallelism, a quantum computer can surpass, in theory, the conventional digital computer in solving certain problems such as factorization \cite{bib:13} and unstructured database search~\cite{bib:14}. An important application of quantum computing is simulating quantum many-body systems \cite{bib:15,bib:16}, for instance, nuclei \cite{bib:17} and molecules \cite{bib:18,bib:19,bib:20}. With Trotterization and quantum phase estimation algorithm, one can simulate the real-time evolution and compute the energy spectrum of a quantum system by using resources, i.e., quantum gates, that scale polynomially with the system size, evolution time and accuracy \cite{bib:21,bib:22}. In recent years, many efforts have been made to discover new quantum algorithms that are feasible for near-term quantum computers \cite{bib:23}. For example, the variational quantum eigensolver (VQE) \cite{bib:24,bib:25,bib:26} enables a quantum computer with a few qubits to calculate the ground-state energy of small molecules.

In this paper, we propose two quantum algorithms suitable for fault-tolerant and near-term quantum computers respectively, to find the optimal effective Hamiltonian of a quantum many-body system. By ``optimal", we mean that the resulting effective Hamiltonian is the one that best describes the original system among a pre-specified set of candidates in a certain subspace.
In the first approach, we use quantum phase estimation (QPE) and Grover's algorithm to search for an effective model with a fault-tolerant quantum computer.
Grover's algorithm can speed up the complexity of the algorithm to $O(\sqrt{N})$, where $N$ denotes the number of total elements in the searching space.
The second method makes use of the variational algorithm. To do so, we transform the problem into searching for the maximum value of the loss function after applying a time evolution operator corresponding to the test Hamiltonian. The form of the loss function will be discussed in Sec. \uppercase\expandafter{\romannumeral2}. We take the transverse-field Ising model as an example and numerically demonstrate that our method can efficiently find the optimal effective Hamiltonian in a set of candidates that best describes the original system in a certain space of effective models. Different from the conventional methods of finding effective models through renormalization group flow or trial and error, here we can simulate the full Hamiltonian and verify the validity of the resulting theories in the full many-body Hilbert space.

This paper is organized as follows.
In Sec. \uppercase\expandafter{\romannumeral2}, we review the definition and significance of the effective Hamiltonian. In Sec. \uppercase\expandafter{\romannumeral3} and Sec. \uppercase\expandafter{\romannumeral4}, we introduce the two methods and present numerical results of the second method. Sec. \uppercase\expandafter{\romannumeral5} is the conclusion of this paper.

\section{The effective Hamiltonian}

There are various definitions of the effective Hamiltonian~\cite{bib:27,bib:28,bib:29}. Usually, the effective Hamiltonian is related to a low-energy subspace. We use $P$ to denote the projection onto the subspace. Ideally, the effective Hamiltonian $H_{eff}$ and the original Hamiltonian $H$ acting on the subspace are related up to a unitary transformation $T$, i.e.,
\begin{equation}
HP = TH_{eff}T^{\dagger}P.
\end{equation}
To verify the above relation with a quantum computer, we introduce a time evolution operator in the form 
\begin{equation} \label{evo}
U(t) = e^{-i(H-TH_{eff}T^{\dagger})t}.
\end{equation}
If $P$ corresponds to an invariant subspace of $H$, i.e.,~$[H,P]=0$, we have 
\begin{equation}\label{eq:Upsi}
\ket{\psi} = U(t) \ket{\psi}
\end{equation}
for all time $t$ and state $\ket{\psi}$ in the subspace. Therefore, we can identify the optimal effective Hamiltonian in a set of candidates by simulating the time-evolution operator and testing Eq.~\ref{eq:Upsi}. In practice, we may take a subspace that is not strictly invariant, such as a truncation on the momentum of single particles. Even in this case, usually, the leakage into states outside the subspace is negligible, and our approach of finding the effective Hamiltonian still works. We will discuss the error introduced by leakage in Section \uppercase\expandafter{\romannumeral3}-C.

In our algorithm, we find the optimal effective Hamiltonian in a set of candidates by maximizing the loss function corresponding to Eq.~\ref{eq:Upsi}. Usually, we need multiple states and evolution times. We use $(\ket{\psi_i}, t_i)$ to denote a trial with an initial state $\ket{\psi_i}$ and evolution time $t_i$. The transition amplitude of one trial is $f_i = \bra{\psi_i}U(t_i)\ket{\psi_i}$, and for our first approach, we can define the loss function of $N_t$ trials as 
\begin{equation}\label{loss1}
F = \left\vert\frac{1}{N_t}\sum_{i=1}^{N_t} f_i\right\vert.
\end{equation}
To evaluate the loss function, we can either measure the transition amplitude of each trial or directly measure the overall transition amplitude by introducing an ancilla system. With the ancilla system, we prepare a composite initial state in the form 
\begin{equation}\label{eq:psi}
\ket{\tilde{\psi}} = \frac{1}{\sqrt{N_t}}\sum_{i=1}^{N_t}\ket{i}\otimes\ket{\psi_i},
\end{equation}
where $\ket{i}$ is the state of an ancilla system. Then, the loss function reads 
\begin{equation}
F = \left\vert \bra{\tilde{\psi}} \tilde{U} \ket{\tilde{\psi}} \right\vert,
\end{equation}
where the composite evolution operator is 
\begin{equation} \label{eq-U}
\tilde{U} = \sum_{i=1}^{N_t} \ket{i}\bra{i}\otimes U(t_i).
\end{equation}
The way to define the loss function is not unique. For our second approach, we can define the loss function as 
\begin{equation}
F_{ave} = \frac{1}{N_t}\sum_{i=1}^{N_t} \abs{f_i}^2,
\end{equation}
and we call it the average fidelity. Regardless of the form, a loss function closer to one indicates a Hamiltonian is a better candidate. In this paper, we find the optimal effective Hamiltonian in a set of candidates $\left\{ H_{eff}(x) \right\}$, where $x$ is the label of the effective Hamiltonian. Although the exact effective Hamiltonian is unknown, usually we have the intuition for which terms may present in the effective Hamiltonian according to the perturbation theory, renormalization theory, or symmetry of the original Hamiltonian~\cite{bib:30,bib:31,bib:32,bib:33}. Given the elementary terms, the problem is to select proper terms and determine their coefficients. The candidate Hamiltonian operators can be generated from these elementary terms. When the effective Hamiltonian is formed of a fixed set of terms and their coefficients are to be determined, $x$ denotes the coefficients. In general, the Hamiltonian set can also include different combinations of terms. For each $x$, there is a corresponding loss function $F(x)$. According to the definitions of the effective Hamiltonian~\cite{bib:27,bib:28,bib:29}, when $TH_{eff}T^{\dagger}$ is an exact effective Hamiltonian, $H_{eff}$ is also the effective Hamiltonian on the subspace. Hence, we can assume that $T$ is identity without loss of generality. The optimal effective Hamiltonian in a set of candidates is $H_{eff}(x^*)$, and $x^* = \arg \max F(x)$. We will give two algorithms for finding the optimal effective Hamiltonian. 

\section{Accelerated search of the effective Hamiltonian}

In this section, we present a quantum algorithm for finding the optimal effective Hamiltonian in a set of candidates. By taking the quantum advantage in unstructured search, we can find the effective Hamiltonian operators with the loss function defined by Eq.~\ref{loss1} above a threshold value in $\sqrt{N}$ evaluations of the loss function, where $N = \abs{\left\{ H_{eff}(x) \right\}}$ is the number of candidate Hamiltonian operators. In this algorithm, quantumness speedups computing in two ways. First, the quantum computer evaluates the loss function defined by Eq.~\ref{loss1} of a many-body system under real-time evolution, which is usually intractable in a classical computer. Second, the unstructured search is accelerated with Grover's algorithm \cite{bib:14} as a subroutine. 

\begin{figure}[!htb]
\centering\includegraphics
[width=1\hsize]
{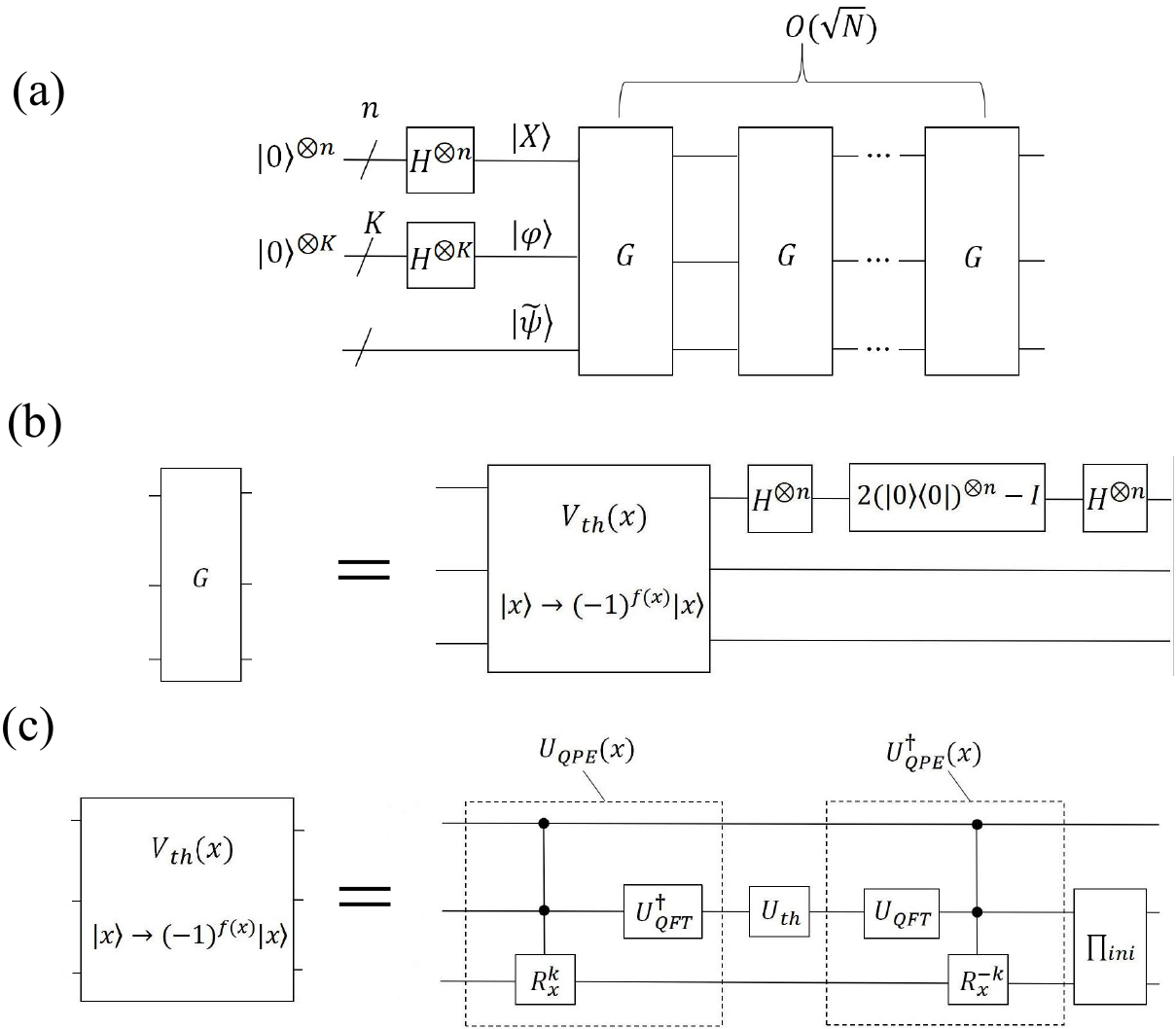}
\caption {(a) The quantum circuit of our algorithm, where the blocks labeled G indicate Grover's search operators. The detailed circuit for G is shown in (b), and the loss function-dependent phase flip operator $V_{th}(x)$ is shown in (c).}
\label{fig:1}
\end{figure}

To implement our algorithm, we need three registers storing the values of $x$, loss function $F(x)$, and state $\ket{\tilde{\psi}}$, respectively, see Fig.~\ref{fig:1}. The algorithm is formed of two subroutines. The first is amplitude amplification and estimation used for evaluating the loss function. The second is Grover's algorithm, in which we find $x$ satisfying the condition $F(x) > \cos\frac{\theta_{th}}{2}$, where $\cos\frac{\theta_{th}}{2}$ is the threshold of the loss function.

\subsection{Loss function evaluation}

We evaluate the loss function $F(x) = \left\vert \bra{\tilde{\psi}} \tilde{U}(x) \ket{\tilde{\psi}} \right\vert$ according to the amplitude amplification and estimation algorithm \cite{bib:34}. As shown in Fig.~\ref{fig:1}(c), this is realized using the quantum phase estimation algorithm~\cite{bib:14} within Grover's search routine. The elementary transformation in amplitude amplification is formed of two phase-flip operators conditioned on the state, which reads 
\begin{equation} \label{eq-Rx}
R_x = \tilde{U}(x)(2\ketbra{\tilde{\psi}}{\tilde{\psi}} - \openone)\tilde{U}(x)^\dag (2\ketbra{\tilde{\psi}}{\tilde{\psi}} - \openone). 
\end{equation}
Here, $\tilde{U}(x)$ is the composite evolution operator corresponding to the effective Hamiltonian $H_{eff}(x)$. Given the initial state $\ket{\tilde{\psi}}$, $R_x$ is equivalent to a rotation in the subspace spanned by $\left\{\ket{\tilde{\psi}}, \tilde{U}(x)\ket{\tilde{\psi}}\right\}$. The rotation is expressed as $R_x = e^{i\sigma_y \theta_x}$, where $\sigma_y$ is a Pauli operator, and we can find $\cos\frac{\theta_x}{2} = F(x) = \left\vert \bra{\tilde{\psi}} \tilde{U}(x) \ket{\tilde{\psi}} \right\vert$ is the loss function. The Pauli operator acts on the two-dimensional subspace, whose eigenstates are $\ket{y\pm}$ corresponding to eigenvalues $\pm 1$, respectively. The relation between eigenstates and $\left\{\ket{\tilde{\psi}}, \tilde{U}(x)\ket{\tilde{\psi}}\right\}$ is given by 
\begin{equation}
\ket{\tilde{\psi}}=\frac{1}{\sqrt{2}}\left(\ket{y_+}+\ket{y_-}\right)
\end{equation}
and
\begin{equation}
R_x\ket{\tilde{\psi}}=\frac{1}{\sqrt{2}}\left(e^{i\theta_x}\ket{y_+}+e^{-i\theta_x}\ket{y_-}\right).
\end{equation}

The loss function is evaluated by reading phases $e^{\pm i\theta_x}$ using the quantum phase estimation algorithm. To implement the algorithm, we need to introduce a register in addition to the register of $\ket{\tilde{\psi}}$, which is initialized in the state $\ket{\varphi} = \frac{1}{\sqrt{K}}\sum_{k=0}^{K-1} \ket{k}$. Here, the integer $K$ determines the resolution of the loss function evaluation. The quantum phase estimation is accomplished with the transformation 
\begin{equation} \label{eq-qpe}
U_{QPE}(x) = \sum_{k=0}^{K-1} \left(U_{QFT}^\dag\ketbra{k}{k}\right)\otimes R_x^k, 
\end{equation}
where $U_{QFT}$ denotes the quantum Fourier transformation \cite{bib:14}. The quantum circuit for $R_x^k$ is shown in Fig.~\ref{fig:Rx} of Appendix A, and the detailed process is summarized below:
\begin{equation} \label{eq9}
	\begin{aligned}
	  &\ket{0}\otimes\ket{\tilde{\psi}}\quad\quad\quad\quad\quad\quad\quad\quad\quad\quad\text{initialize state}\\
	   \longrightarrow&\frac{1}{\sqrt{K}}\sum\limits_{k=0}^{K-1}\ket{k}\otimes\ket{\tilde{\psi}}\quad\quad\quad\quad\quad\quad \text{create superposition}\\
	   \longrightarrow&\frac{1}{\sqrt{K}}\sum\limits_{k=0}^{K-1}\ket{k}\otimes R^k_x \ket{\tilde{\psi}}\quad\quad\quad\quad \ \ \text{apply R gate}\\
	   =&\frac{1}{\sqrt{2K}}\left(\sum\limits_{j=0}^{K-1}e^{ik\theta_x}\ket{k}\otimes\ket{y_+}+\sum\limits_{k=0}^{K-1}e^{-ik\theta_x}\ket{k}\otimes\ket{y_-}\right)\\
	   \longrightarrow&\frac{1}{\sqrt{2}}\left(\ket{A_+}\otimes\ket{y_+}+\ket{A_-}\otimes\ket{y_-}\right) \ \text{inverse $U_{QFT}$.}
	\end{aligned}
\end{equation} 
$\ket{A_+}$ and $\ket{A_-}$ are given by
\begin{eqnarray}
\ket{A_+} &=& \sum_{\theta} f_{\theta}\ket{\theta}, \\
\ket{A_-} &=& \sum_{\theta} f_{\theta}^*\ket{-\theta},
\end{eqnarray}
where $\theta = \frac{2\pi m}{K}$, $m = -\frac{K}{2}+1,\ldots,-1,0,1,\ldots,\frac{K}{2}$ assuming $K$ is even, $\ket{-\pi} \equiv \ket{\pi}$ and $f_\theta = \frac{1}{K}\sum_k e^{i(\theta_x-\theta)k}$. The distribution $\abs{f_\theta}$ is concentrated at $\theta_x$ with the width $\sim \frac{2\pi}{K}$. If $\theta_x$ is in $\{\theta\}$, $\ket{A_\pm} = \ket{\pm \theta_x}$. Therefore, the information of the loss function has been stored in the register $\ket{\theta}$, and we can extract the information by measuring the value of $\theta$. Note that to carry out the accelerated search, here $\ket{A_\pm}$ will not be directly measured. 

\subsection{Loss function-dependent approximate phase flip}

We now introduce the method to amplify amplitudes for $F(x) > \cos\frac{\theta_{th}}{2}$. To do so, we need to realize a phase flip depending on the loss function, i.e., $\theta$. However, because the loss function estimation has a finite resolution, the phase flip is inexact. We approximate the loss function-dependent phase flip with the operator 
\begin{equation}
V_{th}(x) = \Pi_{ini} U_{QPE}(x)^\dag (U_{th}\otimes\openone) U_{QPE}(x),
\end{equation}
where $\Pi_{ini} = \ketbra{\varphi}{\varphi}\otimes\ketbra{\tilde{\psi}}{\tilde{\psi}}$, $U_{th} = \openone -2\sum_{-\theta_{th}< \theta < \theta_{th}}\ketbra{\theta}{\theta}$ is the phase flip depending on $\theta$. 

Because the distribution of $\theta$ is concentrated at $\theta_x$, the $\theta$-dependent phase flip results in a $\theta_x$-dependent phase. Firstly, let's suppose $\theta_x \in \{\theta\}$: If $\abs{\theta_x}< \theta_{th}$, we have $U_{th}\ket{A_\pm} = -\ket{A_\pm}$; If $\abs{\theta_x}\geq \theta_{th}$, we have $U_{th}\ket{A_\pm} = \ket{A_\pm}$. In general $\theta_{th}$ is not one of $\theta$ values because of the finite resolution, and in this case, the effect of $U_{th}$ is not a simple phase change. We note that this problem caused by the finite resolution is only severe when $\theta_x$ is close to $\pm \theta_{th}$. When $\theta_x$ is significantly far away from the threshold $\pm \theta_{th}$, the distribution is concentrated on one side of the threshold, and we can neglect the probability on the other side. 

\begin{figure}[!htb]
\centering\includegraphics
[width=0.9\hsize]
{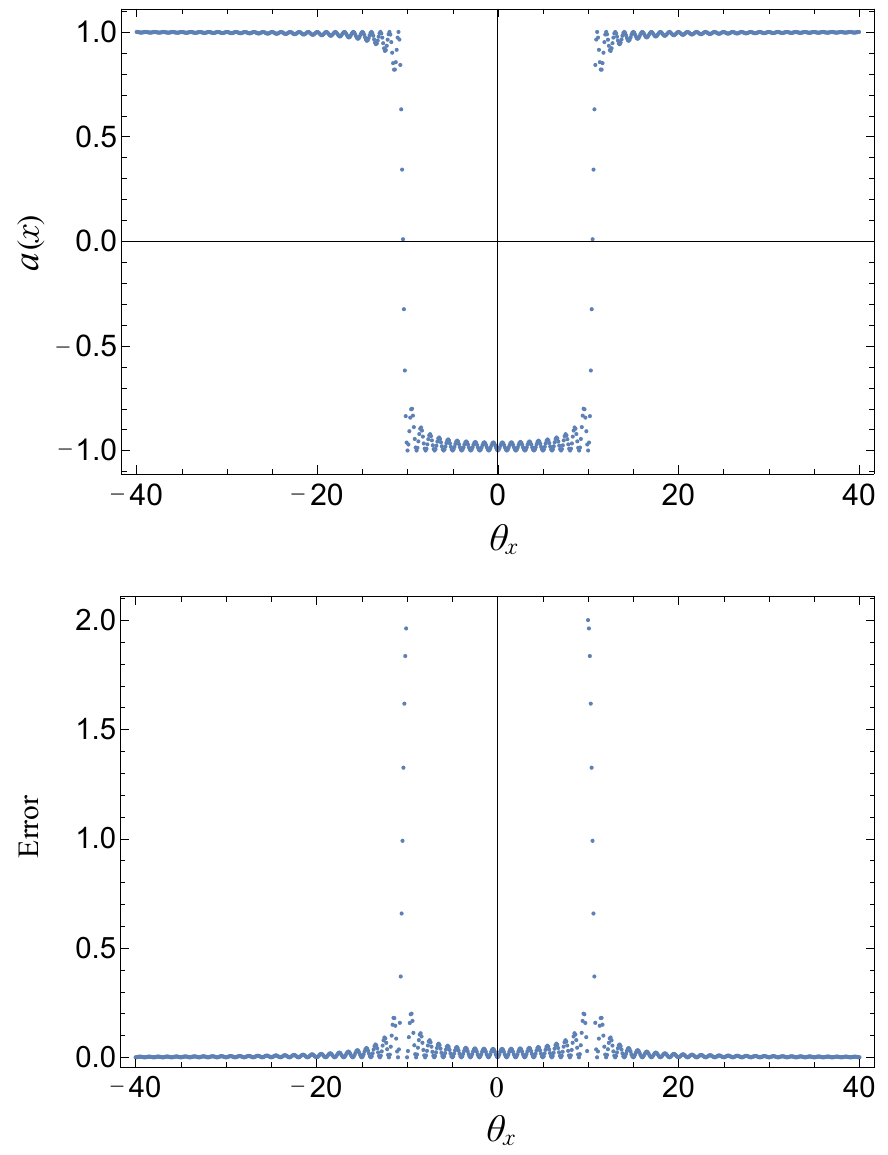}
\caption{Top: the change of $a(x)$ with $\theta_x$. Bottom: the corresponding error $\abs{a(x)-a_x}$. The unit of the Horizontal axis is $\frac{2\pi}{K}$. Here we set $K=4096$ and $\abs{\theta_{th}}=10*\frac{2\pi}{K}$.}
\label{fig:2}
\end{figure}

We can detect the problem caused by the finite resolution by measuring the finial state after an inverse quantum phase estimation operator. If $\theta_x \in \{\theta\}$, a phase is applied on the state by $U_{th}$, and the inverse quantum phase estimation transforms the state back to the initial state $\ket{\varphi}\otimes\ket{\tilde{\psi}}$ up to a $\theta_x$-dependent phase. If $\theta_x \notin \{\theta\}$, the state does not go back to the exact initial state. If the measurement outcome is not $\ket{\varphi}\otimes\ket{\tilde{\psi}}$, we know that the $\theta_x$-dependent phase flip has failed. If the outcome is $\ket{\varphi}\otimes\ket{\tilde{\psi}}$, we obtain the finial state 
\begin{equation}
V_{th}(x)\ket{\varphi}\otimes\ket{\tilde{\psi}} = a(x)\ket{\varphi}\otimes\ket{\tilde{\psi}}, 
\end{equation}
where 
\begin{equation}
a_x = \frac{1}{2}(\bra{A_+}U_{th}\ket{A_+}+\bra{A_-}U_{th}\ket{A_-})
\end{equation}
is always real and in the interval $[-1,1]$. If $\theta_x \in \{\theta\}$, $a_x = \pm 1$ depending on $\theta_x$. If $\theta_x \notin \{\theta\}$ and $\theta_x$ is significantly far away from $\pm \theta_{th}$, we still have $a_x \approx \pm 1$. To illustrate this, we set $K=4096$ and $\abs{\theta_{th}}=10*\frac{2\pi}{K}$, and plot $a(x)$ with a change of $\theta_x$ in Fig.~\ref{fig:2}. We see that the error is significant only when $\theta_x$ is close to $|\theta_{th}|$. With this approximate $\theta_x$-dependent phase flip, we can realize the accelerated search. 

\subsection{Effective-Hamiltonian search with leakage}
We use Grover's algorithm to find effective Hamiltonian operators with a loss function that exceeds the threshold. Accordingly, we need a register to store the parameter $x$, which is initialized in the state $\ket{X} = \frac{1}{\sqrt{N}} \sum_x\ket{x}$. Amplitudes of states with $\abs{\theta_x}< \theta_{th}$ are amplified by iterating two operators $V_X = (\openone - 2\ketbra{X}{X})\otimes\openone\otimes\openone$ and $V_{th}^{ideal}$ on the initial state of the total system $\ket{X}\otimes \ket{\varphi}\otimes \ket{\tilde{\psi}}$. Here, 
\begin{equation}
V_{th}^{ideal} = \sum_x \eta_x\ketbra{x}{x},
\end{equation}
where $\eta_x = -1$ if $\abs{\theta_x}< \theta_{th}$ and $\eta_x = 1$ if $\abs{\theta_x}\geq \theta_{th}$. If the number of effective Hamiltonian operators satisfying $\abs{\theta_x}< \theta_{th}$ is $M$, we can find one of them in $O(\sqrt{N/M})$ iterations \cite{bib:14}. 

In our algorithm, we approximate $V_{th}^{ideal}$ with a controlled $V_{th}(x)$, and the operator on the total system is 
\begin{equation}
V_{th} = \sum_x \ketbra{x}{x}\otimes V_{th}(x).
\end{equation}
Because $\abs{a_x}$ is smaller than one in the vicinity of the threshold $\pm \theta_{th}$, it leads to probability leakage: For each iteration, there is a finite probability that the $\theta_x$-dependent phase flip fails, i.e.,~the measurement outcome is not $\ket{\varphi}\otimes\ket{\tilde{\psi}}$. Next, we analyze the impact of this probability leakage. 

The error due to the difference between $V_{th}$ and $V_{th}^{ideal}$ after $j$ iterations is 
\begin{eqnarray}
&& \normLR{\left[(V_XV_{th}^{ideal})^j-(V_XV_{th})^j\right] \ket{X}} \notag \\
&=& \left\Vert\left[(V_XV_{th}^{ideal})^j-V_XV_{th}(V_XV_{th}^{ideal})^{j-1}\right.\right. \notag \\
&&\left.\left. +V_XV_{th}(V_XV_{th}^{ideal})^{j-1}-(V_XV_{th})^j\right] \ket{X}\right\Vert \notag \\
&\leq & \normLR{\left(V_{th}^{ideal}-V_{th}\right) (V_XV_{th}^{ideal})^{j-1}\ket{X}} \notag \\
&&+ \normLR{\left[(V_XV_{th}^{ideal})^{j-1}-(V_XV_{th})^{j-1}\right] \ket{X}}
\end{eqnarray}
Here, we have neglected $\ket{\varphi}\otimes \ket{\tilde{\psi}}$ for simplicity, we have used that $\norm{V_X} = 1$ and $\norm{V_{th}}\leq 1$ for the matrix 2-norm. Then, the error has the upper bound 
\begin{eqnarray}
&& \normLR{\left[(V_XV_{th}^{ideal})^j-(V_XV_{th})^j\right] \ket{X}} \leq \sum_{i=1}^{j} \epsilon_i,
\end{eqnarray}
where 
\begin{eqnarray}
\epsilon_i = \normLR{\left(V_{th}^{ideal}-V_{th}\right) (V_XV_{th}^{ideal})^{i-1}\ket{X}}.
\end{eqnarray}
Suppose 
\begin{eqnarray}
(V_XV_{th}^{ideal})^{j-1}\ket{X} = \sum_x \alpha_x \ket{x},
\end{eqnarray}
we have 
\begin{eqnarray}
\epsilon_i = \sqrt{\sum_x \abs{\alpha_x}^2 \abs{\eta_x-a_x}}.
\end{eqnarray}
Note that one can work out the expression of $\alpha_x$ straight-forwardly following Grover's algorithm \cite{bib:14}. The error $\abs{\eta_x-a_x}$ is non-zero only in the vicinity of the threshold $\pm \theta_{th}$ with a radius $\sim \frac{2\pi}{K}$, see Fig.~\ref{fig:2}. Therefore, to implement the accelerated effective-Hamiltonian search, we need to choose a sufficiently large $K$ such that the radius is small, and there is no effective Hamiltonian within the radius. 

\subsection{Implementation of the controlled $\tilde{U}(x)$}

In this section, we introduce the implementation details of the operator $\tilde{U}(x)$ which formed the operator $R_x$ in Eq.~\ref{eq-Rx}. Here, we can define the time-evolution operators as
\begin{equation}
U(x,t)=e^{-i(H-H_{eff}(x))t}.
\end{equation}
 $x$ is the label of an effective Hamiltonian, and $t$ is evolution time. According to Eq.~\ref{eq-U}, the operator $\tilde{U}(x)$ is a composite evolution operator
\begin{equation}
 \tilde{U}(x)= \sum_{i=1}^{N_t} \ket{i}\bra{i}\otimes U(x,t_i).   
\end{equation}
For the total system, the operator $U(x,t_i)$ is
\begin{equation} \label{U_X}
 U_X(t_i)=\sum_x \ketbra{x}{x} \otimes U(x,t_i)  
\end{equation}
with different evolution times $t_i$. The following outlines the method for implementing $U_X(t_i)$. Without loss of generality, the candidate effective Hamiltonian can be represented as
\begin{equation}\label{eq-heff}
    H_{eff}(x)=\sum_{i=1}^m\lambda_i H_i.
\end{equation}
Here, the variable $x$ can be expressed as a vector $x=(\lambda_1,\lambda_2,...,\lambda_m)$. $H_i$ represents all possible terms that may appear in the effective Hamiltonian, while $\lambda_i$ denotes their coefficients, which need to be determined by our algorithm, and $m$ is the number of terms. To implement the operator $U_X(t_i)$ , we substitute Eq.~\ref{eq-heff} into Eq.~\ref{U_X} and have
\begin{equation}\label{eq:ux}
\begin{aligned}
U_X(t_i)&=\sum_x \ketbra{x}{x}\otimes e^{-i(H-\sum_{i=1}^m\lambda_i H_i)t_i} \\
&=\sum_x \ketbra{x}{x}\otimes (e^{\frac{-iHt_i}{n_t}}e^{\frac{i\lambda_1 H_1 t_i}{n_t}}...e^{\frac{i\lambda_m H_m t_i}{n_t}})^{n_t} \\
&=\frac{1}{2^{mn/2}} \sum_{i_1,i_2,...,i_m=0}^{i_1,i_2,...,i_m=2^n-1}\ketbra{i_1i_2...i_m}{i_m...i_2i_1} \otimes \\
& \quad (e^{\frac{-iHt_i}{n_t}}(e^{\frac{i \epsilon H_1 t_i}{n_t}})^{i_1}...(e^{\frac{i \epsilon H_m t_i}{n_t}})^{i_m})^{n_t}.
\end{aligned}
\end{equation}
Here, $n_t$ denotes the Trotter steps, and $n$ represents the number of qubits used to encode one coefficient $\lambda_i$. $\epsilon$ is a small constant that signifies the resolution of the coefficients: The second equality in Eq.~\ref{eq:ux} is based on the Trotter-Suzuki decomposition, and
\begin{equation}
\begin{aligned}
&x=(\lambda_1,\lambda_2,...,\lambda_m)=\epsilon*(i_1,i_2 ,...,i_m)\\
&i_m=i_{m,1}2^0+i_{m,2}2^1+...+i_{m,n}2^{n-1}\\
&\ket{i_m}=\ket{i_{m,1}i_{m,2}...i_{m,n}},
\end{aligned}
\end{equation}
where $i_{m,n}=0,1$. For simplicity, we assume that the signs of all terms $\lambda_i H_i$ are contained within $H_i$, and the values of all $\lambda_i$ are positive. It is important to note that we can find some coefficients to be equal to zero in the end, and this indicates that the corresponding terms should not be present in the correct effective Hamiltonian.

In our approach, the number of qubits and quantum gates required for implementing $U_X(t_i)$ increases polynomially with $O(log N)$, where $N$ denotes the number of candidate effective Hamiltonian. As a result, when $N$ increases, our algorithm maintains its ability to achieve quadratic acceleration in the search process, while necessitating additional quantum resources that increase linearly. Note that, the method to implement $U_X(t_i)$ is not unique, and our method described above is relatively universal. The quantum circuit to implement $U_X(t_i)$ can be found in Fig.~\ref{fig:Ux} of Appendix A.

\subsection{Numerical simulations and results}

\begin{figure}[!htb]
\centering\includegraphics
[width=1\linewidth]
{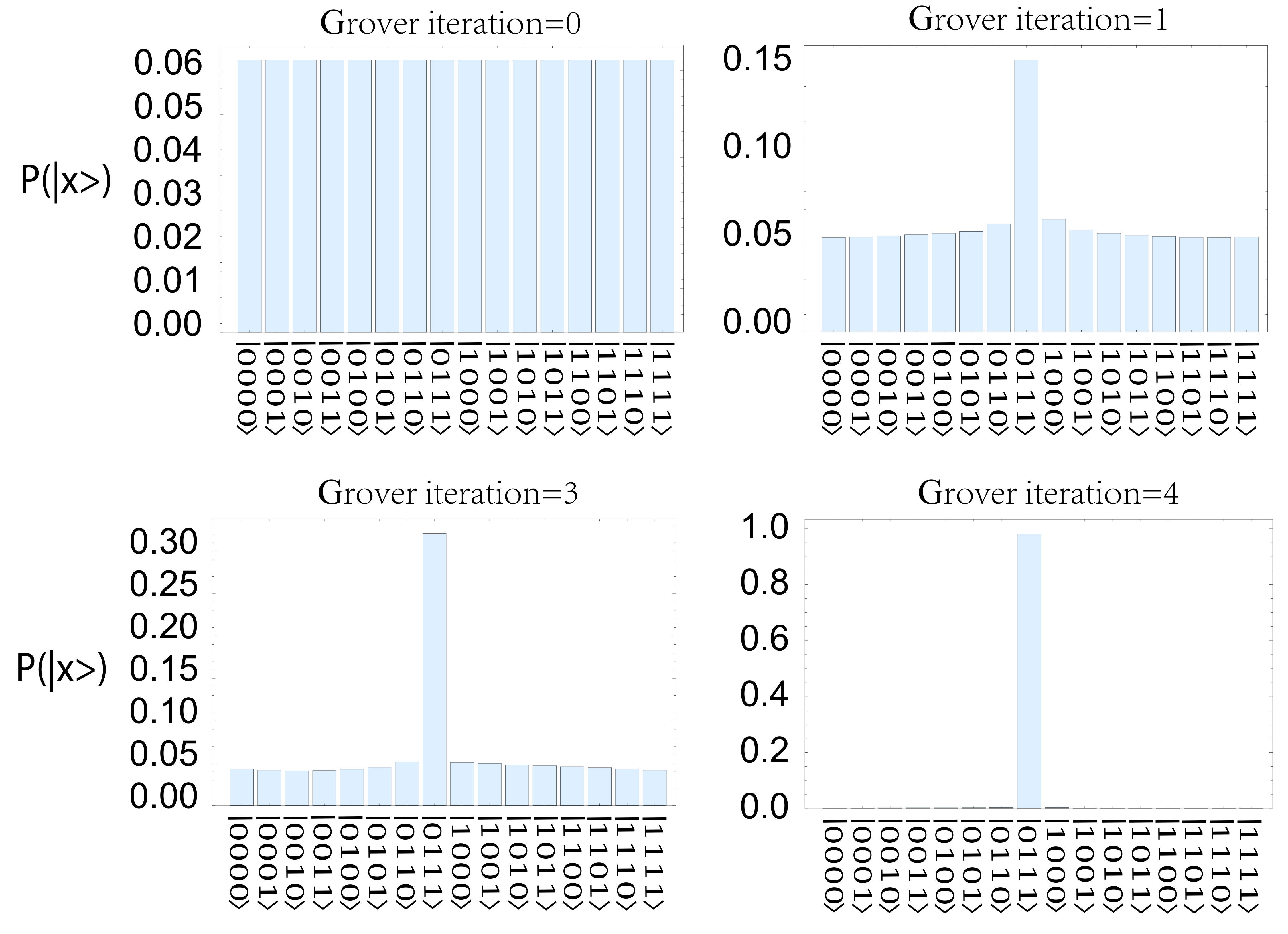}
\caption{The probability distribution P$(\ket{x})$ of the state $\ket{X} = \frac{1}{\sqrt{2^4}} \sum_{x}\ket{x}$ while applying Grover iteration to search for the coefficient of the effective Hamiltonian. The target state is set to be $\ket{0111}$.}
\label{grover}
\end{figure}

To demonstrate the quadratic speedup of our algorithm in searching for the effective Hamiltonian, we present a numerical example in this section. We take the transverse-field Ising model \cite{bib:44} as an example to find its effective Hamiltonian. The Hamiltonian of the whole system is
\begin{equation} \label{eq12} 
	H=-\frac{\Delta}{2}\sum\limits_{i=1}^N\sigma^z_{i}-J\sum\limits_{i=1}^{N-1}\sigma^x_i\sigma^x_{i+1}. \qquad(\Delta\gg J)
\end{equation}
Here, $\sigma^z_{i}$ and $\sigma^x_i$ are Pauli operators acting on the $i$-th site. $N$ denotes the number of spins. $\Delta$ and $J$ are coefficients.
For this system, we can obtain the exact second-order efficient Hamiltonian on a low-energy subspace with the Schrieffer-Wolff (SW) method \cite{bib:45,bib:46,bib:47,bib:48,bib:49,bib:50,bib:51}:
\begin{equation} \label{eq14} 
	\begin{aligned}
H_{eff}&=-\frac{\lambda}{2}\sum\limits_{i=1}^{N-1}(\sigma^x_i\sigma^x_{i+1}+\sigma^y_i\sigma^y_{i+1})\\
&-\frac{\kappa}{2}\sum\limits_{i=1}^{N-2}(\sigma^x_i\sigma^x_{i+2}+\sigma^y_i\sigma^y_{i+2})-\sigma^z_1-\sigma^z_N,
    \end{aligned}	
\end{equation}
where $\lambda$ and $\kappa$ satisfy $\lambda=J$ and $\kappa=\frac{J^2}{2\varDelta}$. Here, the form of the projector is $P=\sum\limits_{i=1}^N\ket{\phi_i}\bra{\phi_i}$, where the vector $\ket{\phi_i}$ denote the n-th single-particle excited state and we can encode the state as $\ket{\phi_1}=\ket{00..001}$ and $\ket{\phi_2}=\ket{00..010}$ and so forth. The form of transformation $T$ is complex and determined by the Schrieffer-Wolff (SW) method. We performed numerical simulations using Grover's algorithm to search for the effective Hamiltonian of a transverse-field Ising model. The process is simulated using Questlink~\cite{bib:52,bib:53} on a classical computer.

The results are shown in Fig.~\ref{grover}. In our simulations, we set $N=2$, $\Delta=10$, and $J=1$. $\epsilon$ is set to be $1/7$. Therefore, the coefficient should satisfy $\lambda=J=7\epsilon$. The evolution time is $t=\pi$, and the Trotter steps is taken to be 1000. We use 10 qubits to implement our algorithm, of which 4 qubits are for encoding the coefficients $x$, 4 qubits are for implementing the quantum Fourier transform in Eq.~\ref{eq-qpe}, and 2 qubits are used for simulating the transverse-field Ising model. We numerically display the probability distribution of the computational basis $\ket{x}$ when applying the Grover iteration for 0, 1, 3, and 4 times, respectively. For a classical algorithm, it requires $2^4$ unstructured searching operations to search for the correct coefficients. However, our algorithm needs only $\sqrt{2^4}$ Grover oracle operations to find the target state with high probability, as shown in Fig.~\ref{grover}.

\section{Variational search of the effective Hamiltonian}
 \begin{figure*}
\centering\includegraphics
[width=0.9\hsize]
{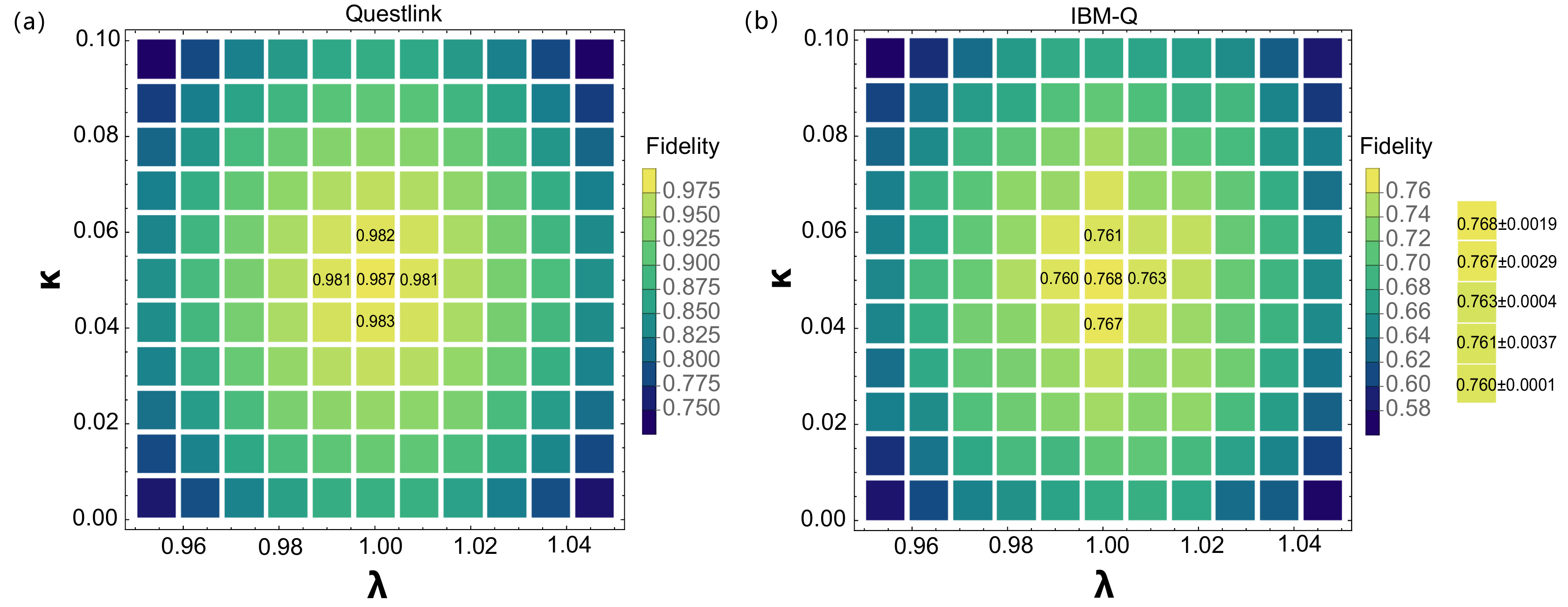}
 \caption{The average fidelity $F_{ave}$ with the change of the coefficients $\lambda$ and $\kappa$. The evolution time is fixed to be $t=2\pi$. (a) and (b) represent results from numerical simulations and experiments on the IBM's quantum device respectively.}
 \label{fig:3}
 \end{figure*} 
With limited quantum resources, searching for the effective Hamiltonian based on QPE and Grover's algorithm is not feasible for noisy intermediate-scale quantum (NISQ) devices \cite{bib:35}. In this section, we introduce an alternative method, variational search, to find the optimal effective Hamiltonian in a set of candidates with a shallow quantum circuit, such that it can be implemented on NISQ devices.

\subsection{Theory of the variational quantum simulation}

The variational quantum simulation \cite{bib:36,bib:37,bib:38,bib:39,bib:40,bib:41,bib:42} is a hybrid classical-quantum algorithm that has various applications in solving many-body problems. With a quantum computer, we first have a trial state $\ket{\Psi(\vec{\theta})}$ that can be prepared by a quantum circuit with a parameterized gate set. The state at time $t$ is represented as $\ket{\Psi(t)}=\ket{\Psi(\vec{\theta}(t))}$ with time-dependent parameters $\vec{\theta}(t)$. The variational 
quantum simulation is used to find the best solution $\vec{\theta}(t+\delta t)$ such that the state evolves from $\ket{\Psi(\vec{\theta}(t))}$ to $\ket{\Psi(\vec{\theta}(t+\delta t))}$ according to Schr\"{o}dinger's equation:
\begin{equation} \label{eq20}
\ket{\Psi(\vec{\theta}(t+\delta t))}\approx\ket{\Psi(\vec{\theta}(t))}-i\delta t H_{test}\ket{\Psi(\vec{\theta}(t))}.
\end{equation}
Here, the test Hamiltonian $H_{test}$ is defined by $H-H_{eff}$. $H$, and $H_{eff}$ denotes the original Hamiltonian and the effective Hamiltonian respectively. To optimize the parameters, we take the method introduced in \cite{bib:43}. The evolution of parameters can be found by solving
\begin{equation} \label{eq21}
	\sum\limits_{j}A_{i,j}\dot{\theta_j}=-i C_i,
\end{equation}
where the matrix elements of $A$ and $C$ are defined by 
\begin{equation} \label{eq22}
\begin{aligned}
A_{i,j}=\frac{\partial \bra{\Psi(\vec{\theta}(t))}}{\partial \theta_i}\frac{\partial \ket{\Psi(\vec{\theta}(t))}}{\partial \theta_j}\\
C_i=\frac{\partial \bra{\Psi(\vec{\theta}(t))}}{\partial \theta_i}H_{test}\ket{\Psi(\vec{\theta}(t))}.
\end{aligned}
\end{equation}
Then, we can iteratively update the parameters under
\begin{equation} \label{eq23}
	\vec{\theta}(t+\delta t)\rightarrow\vec{\theta}(t)+\delta t A^{-1}C.
\end{equation}
The overall flow of the variational quantum simulation is summarized as follows: First, we select initial parameters $\vec{\theta}(0)$ and a small time step $\delta t$. Second, we solve Eq.~\ref{eq21} using the classical computer, in which the matrix $A_{i,j}$ and the vector $C_i$ in the equation are evaluated using the quantum computer. Repeating the second step, we can get the solution that approximates Eq.~\ref{eq20}. 

\subsection{Numerical simulations and results}

We take the transverse-field Ising model showing in Eq.~\ref{eq12} as an example. To find the optimal effective Hamiltonian in a set of candidates, we choose $N_t$ trials $(\ket{\psi_i(\vec{\theta}(0))}, t)$ with initial states $\ket{\psi_i(\vec{\theta}(0))}$ and evolution time $t$ to maximize the average fidelity 
\begin{equation}\label{eq26}
F_{ave}(x)= \frac{1}{N_t}\sum_{i=1}^{N_t} \abs{\braket{\psi_i(\vec{\theta}(t))|\psi_i(\vec{\theta}(0))}}^2
\end{equation}
corresponding to Eq.~\ref{eq:Upsi}, where the parameter $\vec{\theta}(t)$ is determined by Eq.~\ref{eq23} with a fixed time step $\delta_t=2\pi/1000$. The trial state at time $t$ is given by
\begin{equation}
\ket{\psi_i(\vec{\theta}(t))}=e^{-i H_{test} t}\ket{\psi_i(0)}.
\label{eq:trial_state}
\end{equation}
We evaluate the average fidelity corresponding to Eq.~\ref{eq26} with different coefficients of the test Hamiltonian $H_{test}$. When given an initial coefficient $x_0=(\lambda_0, \kappa_0)$, we can use the gradient ascent method \cite{bib:54.1} to maximize the loss function defined in Eq.~\ref{eq26}.
In our example, there are only two parameters that need to be determined. Therefore, we employed the grid search method to identify the optimal parameters.  

We performed numerical simulations using Questlink on a classical computer and experiments on IBM's cloud quantum computer "ibmq-athens". We take $N=5$, $\Delta=10$ and $J=1$. Then the exact values of the coefficients in the effective Hamiltonian $H_{eff}$ are $\lambda=J=1$ and $\kappa=\frac{J^2}{2\varDelta}=0.05$. We choose $\sigma^x_i\ket{00000}$ $(i=1,..,5)$ as the initial states, where $\sigma^x_i$ is the Pauli operator acting on the $i$-th site. The ansatz is chosen as the 3-step trotter circuit.

The results are shown in Fig.~\ref{fig:3}. In both simulations and experiments, we see that when $\lambda$ and $\kappa$ are close to the exact values, the average fidelity $F_{ave}$ becomes higher. Owing to circuit noise, measurement error and shot noise (shots = 8192), the resolution of the coefficients to be determined with experimental results is about 0.01$\sim$0.02 and the fidelity obtained from experiments is generally lower than that from simulations. Nevertheless, in both cases, the fidelity reaches the maximum at the correct value of the coefficients $\lambda=1$ and $\kappa=0.05$.

\begin{figure}[!htb]
\centering\includegraphics[width=0.9\hsize]{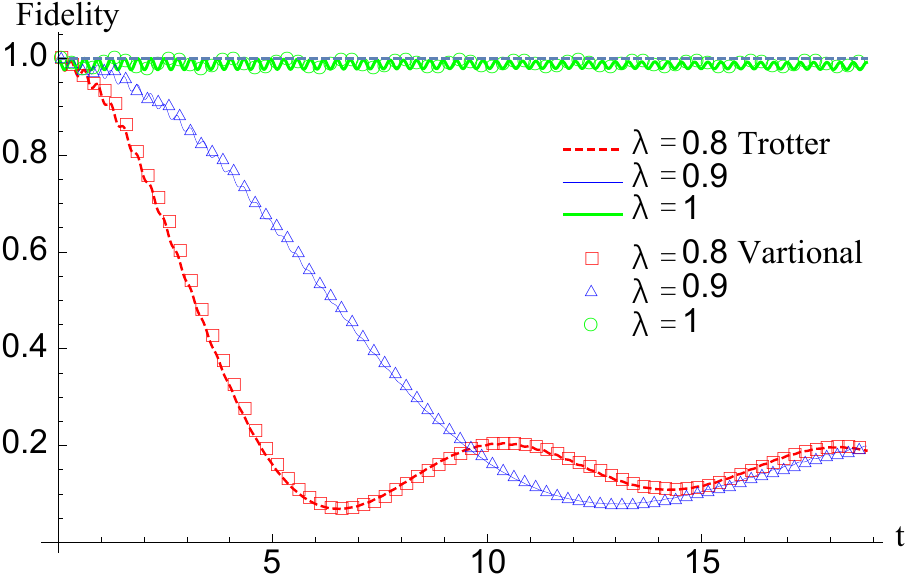}
\caption{The change of average fidelity as the evolution time $t$ increases. The thick green (upper) line, thin blue line, and dashed red line represent results from the trotter method. The green circle, blue triangle, and red square represent results from the variational method. $\kappa$ is fixed to be $0.05$. The average fidelity is evaluated with different $\lambda$.}
\label{fig:6}
\end{figure}

Note that the state in Eq.~\ref{eq:trial_state} can also be realized with the Trotter-Suzuki decomposition~\cite{bib:22} at the cost of deeper quantum circuits. According to the method, the time evolution operator $U(t)$ becomes
\begin{equation} \label{eq19}
	U(t)=e^{-i H_{test} t} \simeq (\prod_{j}e^{-i H_j\tau})^{n_t},
\end{equation}	
where $H_{test}=H-H_{eff}$ is expressed as $H_{test}=\sum_{j} H_j$ and $t$ denotes the total time of evolution. Each term $e^{-i H_j \tau}$ represents the evolution driven by $H_j$ for a short time $\tau$, which can be realized with quantum gates. Usually, $\tau=t/n_t$, and when $n_t$ is larger, the approximation is better. In our simulations we fix $\tau=2\pi/1000$. 

Fig.~\ref{fig:6} shows the evolution of the average fidelity of the quantum states obtained from the Trotter and the variational method simulated classically. We see that the two methods show very similar performance since the two curves nearly overlap as $t$ changes. However, the trotter method uses much deeper circuits, e.g., with over $10^3$ more gates than those required in the variational method at $t=6\pi$. Therefore, our method is more suitable for NISQ devices.

\section{CONCLUSION}
To summarize, we propose two quantum algorithms to search for the effective Hamiltonian of many-body systems. Based on the assumption that the possible terms in the effective Hamiltonian are known but their coefficients are to be determined, our methods can find the optimal effective Hamiltonian in a set of candidates with different coefficients. We define a loss function to measure how close the candidate effective Hamiltonian is to the exact one. when the loss function reaches the maximum value of 1, we find the exact solution. Therefore, the Hamiltonian searching problem is converted to a loss function maximizing problem which the two quantum algorithms are designed for. In the first algorithm, we 
make use of the quantum phase estimation algorithm to encode the loss function into phase information and Grover's algorithm to speed up the complexity of searching to $O(\sqrt{N})$, where $N$ denotes the total number of all candidate effective Hamiltonians $H_{eff}(x)$. Performing this quantum algorithm requires deep circuits that are not feasible for near-term quantum computers. Therefore, we introduce the second quantum algorithm, based on the variational simulation method, which is suitable for NISQ devices. We define a test Hamiltonian $H_{test}=H-H_{eff}$ and realize the time evolution operator $e^{-i t H_{test}}$ with a quantum computer. Both numerical simulations and experiments on an IBM quantum machine were conducted. The results suggest our method can successfully find the optimal effective Hamiltonian in a set of candidates with noise and limited measurement shots. 
Our algorithms show that a quantum computer without fault tolerance can also search for the effective Hamiltonian. Thus, it's an interesting topic for future research, to find other applications of near-term quantum computers for studying the effective theory.

\begin{acknowledgments}
We acknowledge the use of simulation toolkit QuESTlink~\cite{bib:52,bib:53} and IBM Quantum services~\cite{bib:55} for this work. We acknowledge the support of the National Natural Science Foundation of China (Grant No. 11875050, 12088101 and 12275259) and NSAF (Grant No. U1930403).
\end{acknowledgments}
 \begin{appendix}
 \section{Detailed quantum circuits for Grover's algorithm}

 \begin{figure*}[!htb]
\centering\includegraphics
[width=1\linewidth]
{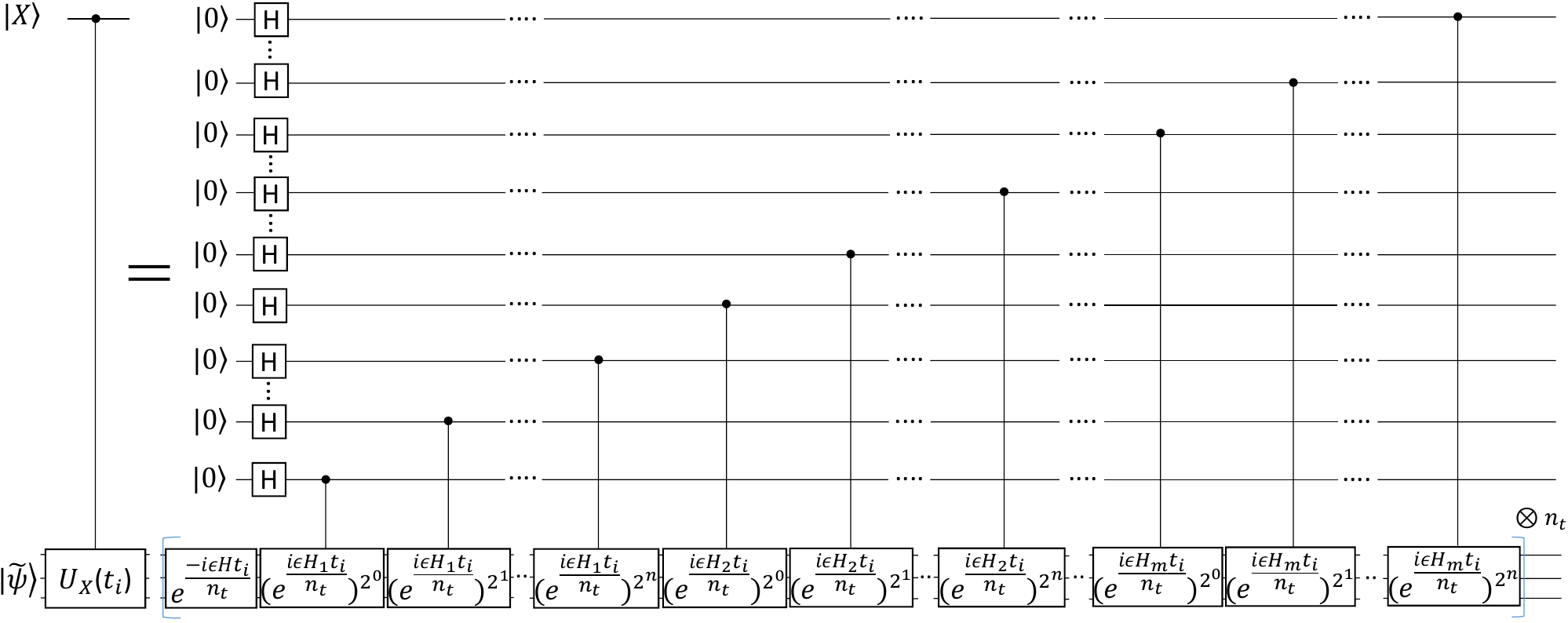}
\caption{The quantum circuit for performing the operator $U_X(t_i)$ in Eq.~\ref{eq:ux}. $n_t$ denotes the trotter steps, and $n$ is the number of qubits used to encode one coefficient $\lambda_i$. $\epsilon$ is a small constant that determines the resolution of the coefficient to be found.}
\label{fig:Ux}
\end{figure*}
 
\begin{figure*}[!htb]
\centering\includegraphics
[width=1\linewidth]
{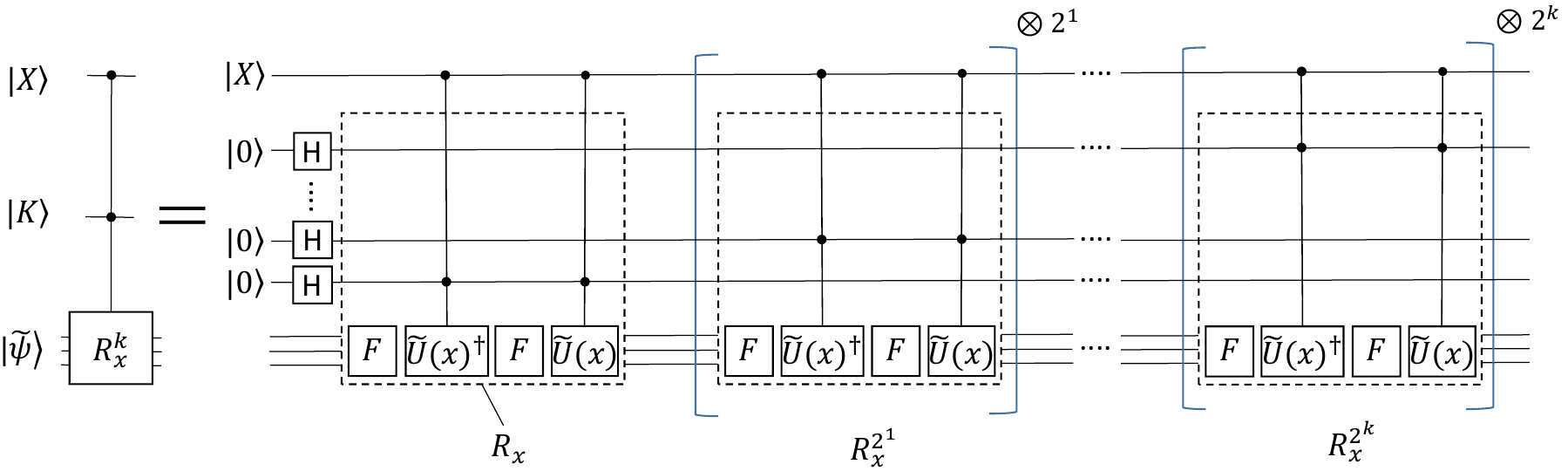}
\caption{The quantum circuit for performing the operator $R_x^k$ in Eq.~\ref{eq-qpe}. The operator $F=2\ketbra{\tilde{\psi}}{\tilde{\psi}} - \openone$.}
\label{fig:Rx}
\end{figure*}

In Fig.~\ref{fig:Ux} and Fig.~\ref{fig:Rx}, we show the detailed quantum circuits for the operator $U_X(t_i)$ and $R_x^k$.
 \end{appendix}

\end{document}